\documentclass[a4paper,11pt,english]{ijamas}

\usepackage[T1]{fontenc}
\usepackage{color}
\usepackage{amsmath}
\usepackage{amssymb}
\usepackage{stackrel}
\makeatletter
\usepackage{tikz}
\usepackage{float}
\usepackage{pgfplots}
 \usetikzlibrary{arrows}

\makeatother

\usepackage{babel}

\begin{document}


\title{Analytic structures of unitary RSOS models with integrable boundary
conditions}

\author{\textbf{Omar El Deeb$^1$ }}

\date{$^1$Department of Physics\\
Modern University of Business and Science\\
Damour, Lebanon\\
odeeb@mubs.edu.lb\\[0.3cm]}

\maketitle


\begin{abstract}
\noindent \emph{In this paper, we consider the unitary critical restricted-solid-on-solid
(RSOS) lattice $\mathcal{M}(5,6)$ model with integrable boundary conditions.
We introduce its commuting double row transfer matrix satisfying the
universal functional relations, and we use it in order to study the
analytic structure of the transfer matrix eigenvalues and plot representative
zero configurations of sample eigenvalues of the transfer matrix.
We finally conclude with a comparative analysis with the critical
and tricritical Ising models with integrable boundary conditions.
}

\medskip


\medskip


\end{abstract}


\section{Introduction}

Integrable models can be solved in finite volumes due to the infinite
number of conservation laws that they have in 1+1 dimensional problems.
The energy spectrum can be fully determined in this case, while it
is a very difficult task in general. The Thermodynamic Bethe Ansatz
(TBA) method allows us to calculate the vacuum polarization effects
of the ground state and its energy. \cite{Zam1,Zam2,Zam21}. Another important and challenging task is to extend this method in order to determine the excited state spectra. The analytic continuation method provides some information about some excited states using the ground state TBA equations as was done in \cite{DoTa}, but this method fails in obtaining the full excitation spectrum for many models including the non-unitary $\mathcal{M}(3,5)$ and the scaling Lee-Yang model \cite{BajDeeb,Lee}.

However, there exists already a powerful and systematic way to obtain the TBA
integral equations for excited states by solving the functional relations
obtained from the Yang-Baxter regularization \cite{PeKlum,KlumPe1,KlumPE,BaxBook}.
Their solutions can be used to fully determine the excitation spectrum
by exploiting analytic and asymptotic properties. This approach was
successfully implemented in solving the tricritical Ising model $\mathcal{M}(4;5)$
with conformal boundary conditions \cite{PCA1,PCA2}. The lattice
regularization approach was also used to solve the Lee-Yang theory
\cite{BPZ,BDeebP,Deeb} as well as the $\mathcal{M}(3,5)$ model \cite{Deeb35}.

In this paper we consider the critical un\textcolor{black}{itary }$\mathcal{M}(5,6)$\textcolor{black}{}
lattice model with integrable boundary conditions. We introduce its
commuting double row transfer matrix satisfying the universal functional
relations, and we use it in order to study the analytic structure
of the transfer matrix eigenvalues and plot representative zero configurations
of sample eigenvalues of the transfer matrix. We finally conclude
with a comparative analysis with related unitary models with
integrable boundary conditions.

The paper is structured as follows: in Section 2,
the conformal model of the $A_{5}$ RSOS lattice model of Forrester-Baxter
\cite{Riggs,ABF,ForBax} is explored in Regime III with crossing
parameter $\lambda=\frac{\pi}{6}$. It introduces the commuting double
row transfer matrices with integrable boundaries. Section 3 analyzes
the conformal spectra of its transfer matrices. We investigate the
analytic structure of the transfer matrix eigenvalues, classify their
excited states in the $(m,n)$ system and plot sample zero configurations
of representative eigenvalues. We then compare the analytic structure and
the zero configuration with corresponding configurations of the related
unitary models like the critical and tricritical Ising models. Section
4 concludes the paper with discussions and future work.


\section{The $\mathcal{M}(5,6)$ Lattice Model}

We analyze the Restricted Solid-on-Solid (RSOS) $\mathcal{M}(5,6)$
lattice model defined on a square lattice built on an
$A_{5}$ Dynkin diagram, with heights differing by $\pm1$ at nearest
neighbor sites. It is one of the $A_{L}$Forrester-Baxter models developed
by \cite{ABF,ForBax,FPR}, with $L=5$ in our case.

The Boltzmann weights of the general $A_{L}$ Forrester-Baxter models
are as follows:

\[
\begin{array}{c}
W\left(\begin{array}{cc}
a\pm1 & a\\
a & a\mp1
\end{array}\right)=\frac{s(\lambda-u)}{s(\lambda)}\end{array}
\]

\begin{equation}
W\left(\begin{array}{cc}
a & a\pm1\\
a\mp1 & a
\end{array}\right)=\frac{g_{a\mp1}}{g_{a\pm1}}\frac{s((a\pm1)\lambda)}{s(a\lambda)}\frac{s(u)}{s(\lambda)}
\end{equation}

\[
W\left(\begin{array}{cc}
a & a\pm1\\
a\pm1 & a
\end{array}\right)=\frac{s(a\lambda\pm u)}{s(a\lambda)}
\]

where $a=1,...,L$ , while $u$ is the spectral parameter. At criticality,
$s(u)=\sin(u)$ and corresponds to the conformal massless model. $\lambda$ is the crossing parameter and it is given by

\begin{equation}
\lambda=\frac{(p'-p)\pi}{p'}
\end{equation}
 where $p'=L+1$ and $p,p'$ are coprime integers with $p<p'$.

The local face weights satisfy the Yang-Baxter equation and this ensures
that the model is integrable. The gauge factors $g_{a}$ are arbitrary
and here they are all set to be equal to $1.$

The critical Forrester-Baxter models in Regime III in the continuum
scaling limit

\begin{equation}
\mbox{Regime III:}\ \ \ \ \ 0<u<\lambda,\ \ \ 0<q<1
\end{equation}
 correspond to the minimal models $\mathcal{M}(p,p')$ whose central
charge is
\begin{equation}
c=1-\frac{6(p-p')^{2}}{pp'}
\end{equation}

In this paper we consider the $\mathcal{M}(5,6)$ model having $\lambda=\frac{\pi}{6}$
and $c=\frac{4}{5}$. A minimal $\mathcal{M}(p,p')$ model has $\frac{(p-1)(p'-1)}{2}$
scaling fields hence the $\mathcal{M}(5,6)$ has ten independent scaling
fields.

\subsection*{Transfer matrices}

The local face weights are used to construct the transfer matrices.
Since the local face weights satisfy the Yang-Baxter equations, we
can show that they form commuting families $[\mathbf{D}(u),\mathbf{D}(v)]=0$. This model satisfies the same functional relation satisfied by the
tricritical hard squares, hard hexagon models and the Lee-Yang model
and the $\mathcal{M}(3,5)$ model \cite{BaxBook,BaxHH,BaxHH1,BaxHS,BDeebP,Deeb,Deeb35}
but with spectral parameter $\lambda=\frac{\pi}{6}$. However, this
model, with its new crossing parameter, has its own analytic structure
with three analyticity strips.

From the Yang-Baxter equations, we can show that the double row transfer
matrices satisfy the functional relation given by

\begin{equation}
\mathbf{D}(u)\mathbf{D}(u+\lambda)=1+\mathbf{Y}.\mathbf{D}(u+3\lambda)\label{eq:YY}
\end{equation}

where $\mathbf{Y}$ in \eqref{eq:YY} is the $\mathbb{Z}_{2}$ height
reversal symmetry.

$E_{n}$, the conformal spectrum of energies of the $\mathcal{M}(5,6)$
model can be obtained through finite size corrections from the logarithm
of the double row transfer matrix eigenvalue. The finite size corrections
in the boundary case are given by

\[
-\log T(u)=Nf_{\mbox{bulk}}(u)+f_{\mbox{boundary}}(u,\xi)+\frac{2\pi}{N}E_{n}\sin\vartheta
\]

where $T(u)$ are the eigenvalues of $\mathbf{D}(u)$, $N$ is the
number of face weights and

\begin{equation}
\vartheta=\frac{\pi u}{\lambda}=6u
\end{equation}
 is the anisotropy angle. $f_{\mbox{bulk }}$ and $f_{\mbox{boundary}}$
are the bulk free energy and the boundary free energy respectively.
$N$ is even in the boundary case.

\subsubsection{Boundary weights}

Commuting row transfer matrices and triangle boundary conditions that
satisfy the left and right boundary Yang Baxter equations guarantee
the integrability of this model. We label the conformal boundary conditions
by the Kac labels $(r,s)$ where $1\leq r\leq4$ and $1\leq s\leq5$.
We limit our study to the $(1,1)$ boundary, as it is a good representative
of the other boundary conditions, with minor differences in their
analytic structures. The $(1,1)$ triangle boundary weights are arbitrary
and they are given by

\begin{equation}
K_{L}\left(\begin{array}{c}
1\\
1
\end{array}2\biggr|u\right)=\frac{s(2\lambda)}{s(\lambda)},\ \ \ \ \ K_{R}\left(2\begin{array}{c}
1\\
1
\end{array}\biggr|u\right)=1
\end{equation}

Other integrable boundary conditions can be constructed by the repeated
action of a seam on the integrable $(1,1)$ boundary \cite{boundaryfusion},
and can be derived automatically. The fact that the boundary weights
satisfy the left and right boundary Yang-Baxter equations ensures
the integrability of the model in presence of those boundaries.

\subsubsection{Double row transfer matrix}

We construct a family of commuting double row transfer matrices $\mathbf{D}(u)$
from the face and triangle boundary weights defined before. For a
lattice of width $N$, transfer matrix $\mathbf{D}(u)$ is given by

\[
\mathbf{D}(u)_{\mathbf{a}}^{\mathbf{b}}=\stackrel[c_{0},..,c_{N}]{}{\sum}K_{L}\Bigl(\begin{array}{c}
r\\
r
\end{array}c_{0}\Bigr|\lambda-u\Bigr)W\left(\begin{array}{cc}
r & b_{1}\\
c_{0} & c_{1}
\end{array}\biggr|\lambda-u\right)W\left(\begin{array}{cc}
b_{1} & b_{2}\\
c_{1} & c_{2}
\end{array}\biggr|\lambda-u\right)...W\left(\begin{array}{cc}
b_{N-1} & s\\
c_{N-1} & c_{N}
\end{array}\biggr|\lambda-u\right)
\]

\begin{equation}
\ \ \ \times W\left(\begin{array}{cc}
c_{0} & c_{1}\\
r & a_{1}
\end{array}\biggr|u\right)W\left(\begin{array}{cc}
c_{1} & c_{2}\\
a_{1} & a_{2}
\end{array}\biggr|u\right)....W\left(\begin{array}{cc}
c_{N-1} & c_{N}\\
a_{N-1} & s
\end{array}\biggr|u\right)K_{R}\left(c_{N}\begin{array}{c}
s\\
s
\end{array}\biggr|u\right)
\end{equation}

It satisfies periodicity $\mathbf{D}(u+\pi)=\mathbf{D}(u)$, commutativity
$[\mathbf{D}(u),\mathbf{D}(v)]=0$ and the crossing symmetry property
$\mbox{\ensuremath{\mathbf{D}}}(u)=\mathbf{D}(\lambda-u)$. In the
general case, $\mathbf{D}(u)$ is not symmetric or normal, but it
can be diagonalized because $\tilde{\mathbf{D}}(u)=\mathbf{G}\mathbf{D(u)}=\tilde{\mathbf{D}}(u)^{T}$
is symmetric where the diagonal matrix $\mathbf{G}$ is given by

\begin{equation}
\mathbf{G}_{\mathbf{a}}^{\mathbf{b}}=\prod_{j=1}^{N-1}G(a_{j},a_{j+1})\delta(a_{j},b_{j})\ \ \ \ \ \mbox{with\ \ \ \ \ \ensuremath{G(a,b)=\begin{cases}
\begin{array}{cc}
\frac{s(\lambda)}{s(2\lambda)}, & \ \ b=1,4\\
1 & \mbox{otherwise}
\end{array}\end{cases}}}
\end{equation}

We introduce the normalized transfer matrix

\begin{equation}
\mathbf{D}(u)=S_{b}(u)\frac{s^{2}(2u-\lambda)}{s(2u+\lambda)s(2u-3\lambda)}\left(\frac{s(\lambda)s(u+2\lambda)}{s(u+\lambda)s(u+3\lambda)}\right)^{N}\mathbf{T}(u)
\end{equation}
In the following analysis we discuss $(1,1)$ left and right boundary
weights corresponding to the $(r,s)=(1,1)$ boundary. The eigenvalues of the normalized double row transfer matrix $\mathbf{T}(u)$
satisfy the functional equation

\begin{equation}
t(u)t(u+\lambda)=1+t(u+3\lambda)
\end{equation}


\section{Conformal Spectra}

In this section, we analyze the complex zero distributions of the
eigenvalues of the double row transfer matrix with emphasis on the
behavior of finite excitations above the ground state.

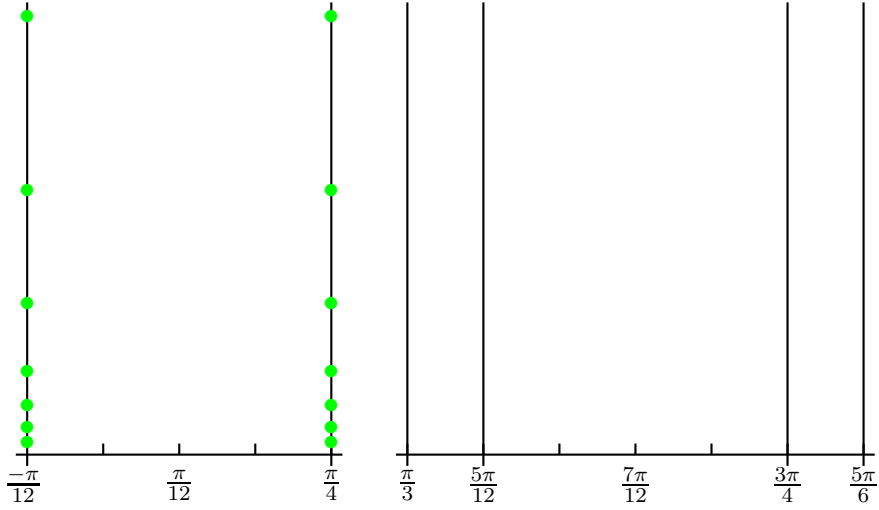
\begin{figure}[H] \begin{center}\begin{tikzpicture}[->,>=stealth',shorten >=0.1 pt,auto,node distance=1cm,   thick,main node/.style={circle,fill=blue!10,draw,font=\sffamily\Large\bfseries}]

\draw[thick,-] (-0.15,0) -- (4.15,0);
\draw[thick,-] (4.85,0) -- (11.15,0);
\draw[thick,-] (0,-0.15) -- (0,6) ;
\draw[thick,-] (4,-0.15) -- (4,6);
\draw[thick,-] (5,-0.15) -- (5,6);
\draw[thick,-] (6,-0.15) -- (6,6);
\draw[thick,-] (10,-0.15) -- (10,6);
\draw[thick,-] (11,-0.15) -- (11,6);

\draw[thick,-] (1,0) -- (1,0.15); \draw[thick,-] (2,0) -- (2,0.15); \draw[thick,-] (3,0) -- (3,0.15); \draw[thick,-] (4,0) -- (4,0.15);  \draw[thick,-] (5,0) -- (5,0.15); \draw[thick,-] (6,0) -- (6,0.15); \draw[thick,-] (7,0) -- (7,0.15); \draw[thick,-] (8,0) -- (8,0.15); \draw[thick,-] (9,0) -- (9,0.15); \draw[thick,-] (10,0) -- (10,0.15);  \draw[thick,-] (11,0) -- (11,0.15);

\node [green] at (0,0.15) {\large \textbullet}; \node [green] at (4,0.15) {\large \textbullet};
\node [green] at (0,0.35) {\large \textbullet}; \node [green] at (4,0.35) {\large \textbullet};
\node [green] at (0,0.65) {\large \textbullet}; \node [green] at (4,0.65) {\large \textbullet};
\node [green] at (0,1.1) {\large \textbullet}; \node [green] at (4,1.1) {\large \textbullet};
\node [green] at (0,2) {\large \textbullet}; \node [green] at (4,2) {\large \textbullet};
\node [green] at (0,3.5) {\large \textbullet}; \node [green] at (4,3.5) {\large \textbullet};
\node [green] at (0,5.8) {\large \textbullet}; \node [green] at (4,5.8) {\large \textbullet};

\node [black] at (-0.05,-0.4) { $\frac{-\pi}{12}$ }; \node [black] at (2,-0.4) { $\frac{\pi}{12}$ }; \node [black] at (4,-0.4) { $\frac{\pi}{4}$ };  \node [black] at (5,-0.4) { $\frac{\pi}{3}$ };\node [black] at (6,-0.4) { $\frac{5\pi}{12}$ };\node [black] at (8,-0.4) { $\frac{7\pi}{12}$ }; \node [black] at (10,-0.4) { $\frac{3\pi}{4}$ }; \node [black] at (11,-0.4) { $\frac{5\pi}{6}$ };

  \end{tikzpicture}
\caption{The zero configuration of the eigenvalue of the transfer matrix corresponding to the ground state. All the zeros are distributed as 2-strings in the first analyticity strip.}
\end{center} \end{figure}

\subsection*{$(m,n)$ systems and zero patterns}

This lattice model corresponds to the conformal field theory model
with central charge $c=\frac{4}{5}$ . The face weights and the triangle
boundary weights are expressed in terms of the trigonometric functions
$s(u)=\sin(u)$. We characterize its eigenvalues by
the locations and patterns of the zeros in the complex $u-$ plane.
The elements of the unrenormalized transfer matrix are Laurent polynomials
in the variables $z=e^{iu}$ and $z^{-1}=e^{-iu}$. The transfer matrices
are commuting families with a common set of $u$-independent eigenvectors.
Consequently, the eigenvalues are also Laurent polynomials of the
same degree. We numerically diagonalize those eigenvalues and obtain
their zeros. They are characterized by the
location and the pattern of the zeros in the complex $u$-plane that are analyzed in terms of the $(m,n)$ systems.

In the boundary case, it is enough to study the eigenvalue zero distributions
on the upper half plane as the transfer matrix is symmetric under
complex conjugation. The zeros form strings and the excitations are
described by their string content in the analyticity strips. In this
paper we only consider the boundary case with $(r,s)=(1,1)$. There are three different analyticity strips in the complex $u$-plane
but the third is a subset of the second. They are given by

\textcolor{black}{
\begin{equation}
\frac{-\pi}{12}<\mbox{Re \ensuremath{u}<\ensuremath{\frac{\pi}{4}}\ ,\ \ }\frac{\pi}{3}<\mbox{Re \ensuremath{u}<\ensuremath{\frac{5\pi}{6}}}\ ,\ \ \frac{5\pi}{12}<\mbox{Re \ensuremath{u}<\ensuremath{\frac{3\pi}{4}}}
\end{equation}
}

In terms of $\lambda$, the analyticity strips in the complex $u-$plane could be written as:

\textcolor{black}{
\begin{equation}
\frac{-\lambda}{2}<\mbox{Re \ensuremath{u}<\ensuremath{\frac{3\lambda}{2}}\ ,\ \ }2\lambda<\mbox{Re \ensuremath{u}<\ensuremath{5\lambda\ ,\ \ \frac{5\lambda}{2}<\mbox{Re \ensuremath{u}<\ensuremath{\frac{\ensuremath{9\lambda}}{2}}}}}
\end{equation}
}
\begin{center}

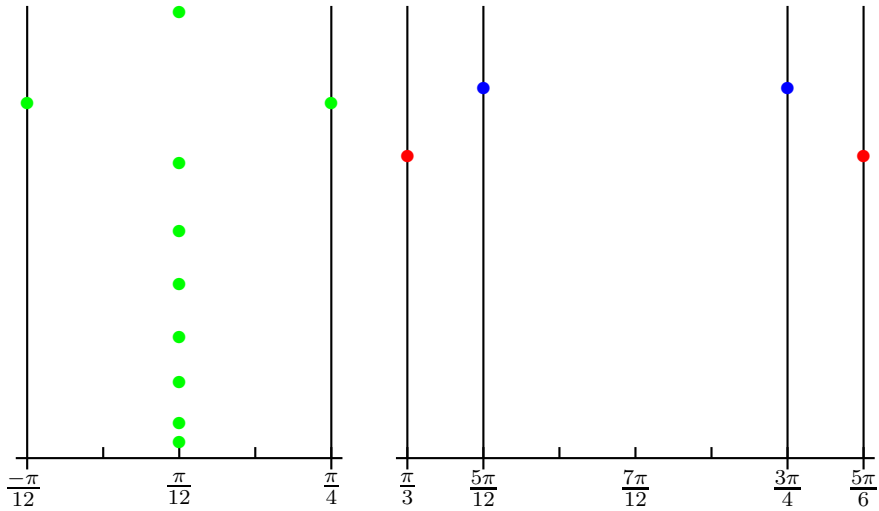
\begin{figure}[H] \begin{center}\begin{tikzpicture}[->,>=stealth',shorten >=0.1 pt,auto,node distance=1cm,   thick,main node/.style={circle,fill=blue!10,draw,font=\sffamily\Large\bfseries}]

\draw[thick,-] (-0.15,0) -- (4.15,0);
\draw[thick,-] (4.85,0) -- (11.15,0);
\draw[thick,-] (0,-0.15) -- (0,6) ;
\draw[thick,-] (4,-0.15) -- (4,6);
\draw[thick,-] (5,-0.15) -- (5,6);
\draw[thick,-] (6,-0.15) -- (6,6);
\draw[thick,-] (10,-0.15) -- (10,6);
\draw[thick,-] (11,-0.15) -- (11,6);

\draw[thick,-] (1,0) -- (1,0.15); \draw[thick,-] (2,0) -- (2,0.15); \draw[thick,-] (3,0) -- (3,0.15); \draw[thick,-] (4,0) -- (4,0.15);  \draw[thick,-] (5,0) -- (5,0.15); \draw[thick,-] (6,0) -- (6,0.15); \draw[thick,-] (7,0) -- (7,0.15); \draw[thick,-] (8,0) -- (8,0.15); \draw[thick,-] (9,0) -- (9,0.15); \draw[thick,-] (10,0) -- (10,0.15);  \draw[thick,-] (11,0) -- (11,0.15);

\node [green] at (0,4.7) {\large \textbullet}; \node [green] at (4,4.7) {\large \textbullet};
\node [green] at (2,0.2) {\large \textbullet}; \node [green] at (2,0.45) {\large \textbullet};\node [green] at (2,1) {\large \textbullet};\node [green] at (2,1.6) {\large \textbullet};\node [green] at (2,2.3) {\large \textbullet};\node [green] at (2,3) {\large \textbullet};\node [green] at (2,3.9) {\large \textbullet};\node [green] at (2,5.9) {\large \textbullet};

\node [red] at (5,4) {\large \textbullet};  \node [red] at (11,4) {\large \textbullet};

\node [blue] at (6,4.9) {\large \textbullet}; \node [blue] at (10,4.9) {\large \textbullet};

\node [black] at (-0.05,-0.4) { $\frac{-\pi}{12}$ }; \node [black] at (2,-0.4) { $\frac{\pi}{12}$ }; \node [black] at (4,-0.4) { $\frac{\pi}{4}$ };  \node [black] at (5,-0.4) { $\frac{\pi}{3}$ };\node [black] at (6,-0.4) { $\frac{5\pi}{12}$ };\node [black] at (8,-0.4) { $\frac{7\pi}{12}$ }; \node [black] at (10,-0.4) { $\frac{3\pi}{4}$ }; \node [black] at (11,-0.4) { $\frac{5\pi}{6}$ };

  \end{tikzpicture}
\caption{A typical configuration of zeros of an eigenvalue of the transfer matrix corresponding to an excited state. The zeros of the first strip are in green, the second in red, and the third in blue.}
\end{center} \end{figure}
\par\end{center}

We notice the occurrence of zeros in all analyticity strips. In the
first strip we assign those patterns as ``1-strings'' and ``2-strings''
formed by single zeroes and pairs of zeroes respectively. In the second, only ``2-strings'' appear while in the third we obtain
again ``1-strings'' and ``2-strings''. The second and the third
strips could be treated as one analyticity strip with a pattern of
long and short 2-strings. However, we follow here the general classification
of RSOS models with more than one analyticity strips for unitary $\mathcal{M}(L,L+1)$
models. Figure 1 gives the zero configuration content for the ground
state eigenvalue of the boundary $\mathcal{M}(5,6)$ model while figures
2 and 3 display sample configurations for eigenvalues corresponding
to excited states .
\begin{center}

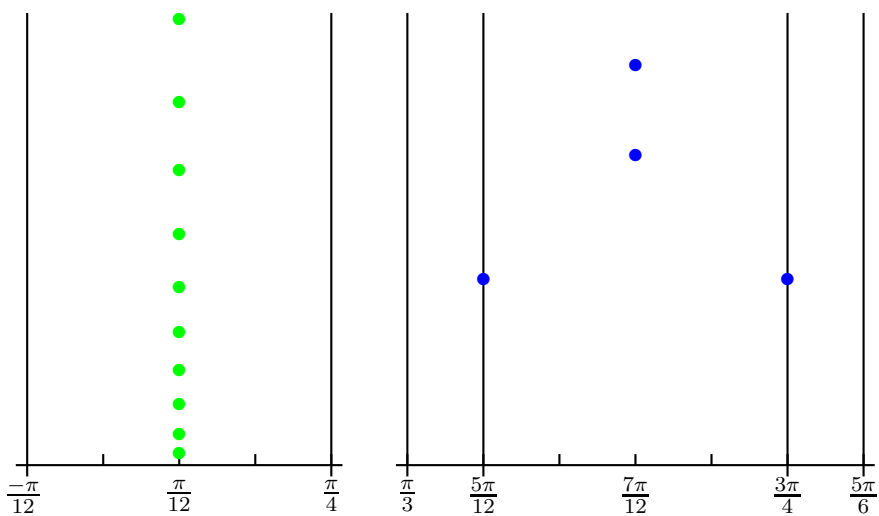
\begin{figure}[H] \begin{center}
\begin{tikzpicture}[->,>=stealth',shorten >=0.1 pt,scale=1,auto=left,node distance=1cm,thick,main node/.style={circle,fill=blue!10,draw,font=\sffamily\Large\bfseries}]

\draw[thick,-] (-0.15,0) -- (4.15,0);
\draw[thick,-] (4.85,0) -- (11.15,0);
\draw[thick,-] (0,-0.15) -- (0,6) ;
\draw[thick,-] (4,-0.15) -- (4,6);
\draw[thick,-] (5,-0.15) -- (5,6);
\draw[thick,-] (6,-0.15) -- (6,6);
\draw[thick,-] (10,-0.15) -- (10,6);
\draw[thick,-] (11,-0.15) -- (11,6);

\draw[thick,-] (1,0) -- (1,0.15); \draw[thick,-] (2,0) -- (2,0.15); \draw[thick,-] (3,0) -- (3,0.15); \draw[thick,-] (4,0) -- (4,0.15);  \draw[thick,-] (5,0) -- (5,0.15); \draw[thick,-] (6,0) -- (6,0.15); \draw[thick,-] (7,0) -- (7,0.15); \draw[thick,-] (8,0) -- (8,0.15); \draw[thick,-] (9,0) -- (9,0.15); \draw[thick,-] (10,0) -- (10,0.15);  \draw[thick,-] (11,0) -- (11,0.15);

\node [green] at (2,0.15) {\large \textbullet}; \node [green] at (2,0.4) {\large \textbullet};\node [green] at (2,0.8) {\large \textbullet};\node [green] at (2,1.25) {\large \textbullet};\node [green] at (2,1.75) {\large \textbullet};\node [green] at (2,2.35) {\large \textbullet};\node [green] at (2,3.05) {\large \textbullet};\node [green] at (2,3.9) {\large \textbullet};\node [green] at (2,4.8) {\large \textbullet}; \node [green] at (2,5.9) {\large \textbullet};

\node [blue] at (8,4.1) {\large \textbullet};  \node [blue] at (8,5.3) {\large \textbullet};

\node [blue] at (6,2.45) {\large \textbullet}; \node [blue] at (10,2.45) {\large \textbullet};

\node [black] at (-0.05,-0.4) { $\frac{-\pi}{12}$ }; \node [black] at (2,-0.4) { $\frac{\pi}{12}$ }; \node [black] at (4,-0.4) { $\frac{\pi}{4}$ };  \node [black] at (5,-0.4) { $\frac{\pi}{3}$ };\node [black] at (6,-0.4) { $\frac{5\pi}{12}$ };\node [black] at (8,-0.4) { $\frac{7\pi}{12}$ }; \node [black] at (10,-0.4) { $\frac{3\pi}{4}$ }; \node [black] at (11,-0.4) { $\frac{5\pi}{6}$ };

  \end{tikzpicture}

\caption{Another configuration of zeros of an eigenvalue of the transfer matrix corresponding to an excited state. Here we see the 1-strings of the first analyticity strip in green together with 1-strings and 2-strings of the third analyticity strips in blue.}

\end{center} \end{figure}
\par\end{center}

In the first strip, a 1-string $u_{j}=\frac{\pi}{12}+iv_{j}$ whose
real part is $\frac{\pi}{12}$ lies in the middle of the analyticity
strip. Each 2-string consists of a pair of zeros whose real parts
are at $\frac{-\pi}{12}$ and $\frac{\pi}{4}$, with equal imaginary
parts, thus $u_{j}=\frac{-\pi}{12}+iy_{j},\frac{\pi}{4}+iy_{j}$.
In the second strip, the 2-string lies at $u_{j}=\frac{\pi}{3}+iy_{j},\frac{5\pi}{6}+iy_{j}$
with equal imaginary parts and with real parts $\frac{\pi}{3}$ and
$\frac{5\pi}{6}$. Finally, the third strip contains a pattern of
a 1-string occurring at $u_{j}=\frac{7\pi}{12}+iv_{j}$ whose real
part is $\frac{7\pi}{12}$ and 2-strings occurring at $u_{j}=\frac{5\pi}{12}+iy_{j},\frac{3\pi}{4}+iy_{j}$with
real parts $\frac{5\pi}{12}$ and $\frac{3\pi}{4}$.

The string contents are described by $(m,n)$ systems \cite{Berkovich}
. The $(1,1)$ sector, in unitary minimal models $\mathcal{M}(L,L-1)$
satisfies the relation:

\begin{equation}
\boldsymbol{m}+\boldsymbol{n}=\frac{1}{2}(N\boldsymbol{e}_{\boldsymbol{1}}+A\boldsymbol{m})
\end{equation}

where $A$ is the adjacency matrix of the $A_{L-2}$ model, $\boldsymbol{e_{1}}=(1,0,..,0)$
, $m=(m_{1},m_{2},...,m_{L-2})$, and $n=(n_{1},n_{2},...,n_{L-2})$.

For the unitary $\mathcal{M}(5,6)$ model, we obtain the relations

\[
m_{1}+n_{1}=\frac{N+m_{2}}{2}\ ,\ \ \ \ m_{2}+n_{2}=\frac{m_{1}+m_{3}}{2}\ ,\ \ \ \ m_{3}+n_{3}=\frac{m_{2}}{2}
\]

where $m$ is the number of short 2-strings, $n$ is the number of
long 2-strings and $N$ is even. We can verify that the number of
zeros $N=2n_{1}+2n_{2}+2n_{3}+m_{1}+m_{3}$ hence the 1-strings corresponding
to the second analyticity strip do not contribute to the zero configuration
plot.

In all of the sectors, the 1-string contributes to one zero. In addition,
2-string contributes two zeroes. Hence, the $(m,n)$ system expresses
the conservation of the $2N$ zeroes in the periodicity strip. The
ground state occurs when all zeros occur as 2-strings in the first
sector solely. The appearance of 1-strings in this sector and all
other strings in the other sectors expresses excited states. The first
excited states are expressed by the 1-strings of the first sector
and 2-strings from the second strip. The appearance of zero patterns
from the 1-string and 2-string content of the third strip represents
the next higher excited states. Note that $n\rightarrow N$ as $N\rightarrow\infty$
while $m_{i}$ and $n_{2}$ and $n_{3}$ are finite for finite excitations.

\paragraph*{Other unitary models}

Several other unitary models were analyzed including the $\mathcal{M}(3,4)$
critical Ising model and the $\mathcal{M}(4,5)$ tricritical Ising
model. We can notice that the analytic structure consists of a single
strip for the critical Ising model with the real part given by \textcolor{black}{
\[
\frac{-\pi}{8}<\mbox{Re \ensuremath{u}<\ensuremath{\frac{3\pi}{8}}}
\]
}

\textcolor{black}{In terms of the spectral parameter $\lambda$,
\[
\frac{-\lambda}{2}<\mbox{Re \ensuremath{u}<\ensuremath{\frac{3\lambda}{2}}}
\]
}

and symmetric with respect to Re $u=\frac{3\pi}{8}$ with $\lambda=\frac{\pi}{4}$.
\begin{center}

\begin{figure}[H] \begin{center}
\begin{tikzpicture}[->,>=stealth',shorten >=0.1 pt,scale=1,auto=left,node distance=1cm,thick,main node/.style={circle,fill=blue!10,draw,font=\sffamily\Large\bfseries}]

\draw[thick,-] (-0.15,0) -- (4.15,0);
\draw[thick,-] (0,-0.15) -- (0,5) ;
\draw[thick,-] (4,-0.15) -- (4,5);

\draw[thick,-] (1,0) -- (1,0.15); \draw[thick,-] (2,0) -- (2,0.15); \draw[thick,-] (3,0) -- (3,0.15);

\node [green] at (2,2.3) {\large \textbullet};\node [green] at (2,3.2) {\large \textbullet};

\node [green] at (0,0.4) {\large \textbullet};\node [green] at (4,0.4) {\large \textbullet};
\node [green] at (0,0.9) {\large \textbullet};\node [green] at (4,0.9) {\large \textbullet};
\node [green] at (0,1.5) {\large \textbullet};\node [green] at (4,1.5) {\large \textbullet};
\node [green] at (0,4.7) {\large \textbullet};\node [green] at (4,4.7) {\large \textbullet};

\node [black] at (-0.05,-0.4) { $\frac{-\pi}{8}$ }; \node [black] at (2,-0.4) { $\frac{\pi}{8}$ }; \node [black] at (4,-0.4) { $\frac{3\pi}{8}$ };

\end{tikzpicture}

\caption{A configuration of zeros of an eigenvalue of the transfer matrix corresponding to an excited state of the Ising model.}

\end{center} \end{figure}
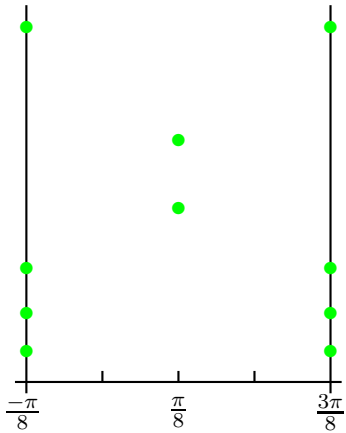
\par\end{center}

The tricritical Ising model consists of two analyticity strips given
by

\[
\frac{-\pi}{10}<\mbox{Re \ensuremath{u}<\ensuremath{\frac{3\pi}{10}}\ ,\ \ }\frac{2\pi}{5}<\mbox{Re \ensuremath{u}<\ensuremath{\frac{4\pi}{5}}}
\]

\textcolor{black}{corresponding to}
\[
\frac{-\lambda}{2}<\mbox{Re \ensuremath{u}<\ensuremath{\frac{3\lambda}{2}}\ ,\ \ }2\lambda<\mbox{Re \ensuremath{u}<\ensuremath{4\lambda}}
\]
with $\lambda=\frac{\pi}{5}$. Sample configurations of zeroes of
eigenvalues representing excited states of those models are given
in figure 4 and figure 5.

In this respect, the $\mathcal{M}(5,6)$ model has a similar structure
to those unitary models, with three analyticity strips as discussed
before. This is a general feature of the $\mathcal{M}(L,L+1)$ unitary
models with $\lambda=\frac{\pi}{L+1}$ with $L-2$ analyiticty strips.
\begin{center}

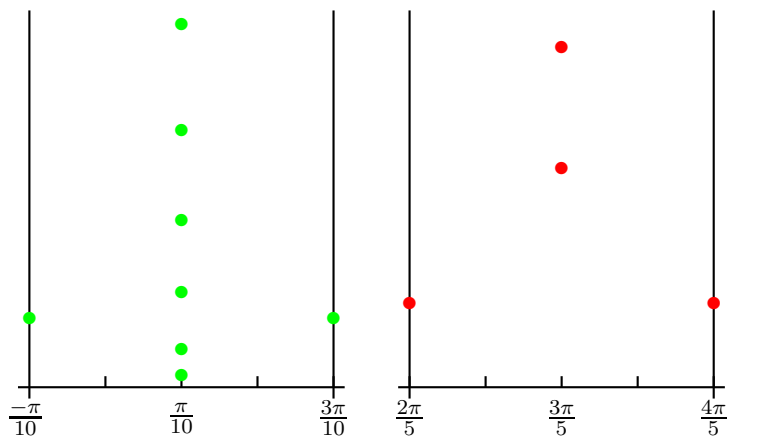
\begin{figure}[H] \begin{center}
\begin{tikzpicture}[->,>=stealth',shorten >=0.1 pt,scale=1,auto=left,node distance=1cm,thick,main node/.style={circle,fill=blue!10,draw,font=\sffamily\Large\bfseries}]

\draw[thick,-] (-0.15,0) -- (4.15,0);
\draw[thick,-] (4.85,0) -- (9.15,0);
\draw[thick,-] (0,-0.15) -- (0,5) ;
\draw[thick,-] (4,-0.15) -- (4,5);
\draw[thick,-] (5,-0.15) -- (5,5);
\draw[thick,-] (9,-0.15) -- (9,5);

\draw[thick,-] (1,0) -- (1,0.15); \draw[thick,-] (2,0) -- (2,0.15); \draw[thick,-] (3,0) -- (3,0.15); \draw[thick,-] (4,0) -- (4,0.15);  \draw[thick,-] (5,0) -- (5,0.15); \draw[thick,-] (6,0) -- (6,0.15); \draw[thick,-] (7,0) -- (7,0.15); \draw[thick,-] (8,0) -- (8,0.15); \draw[thick,-] (9,0) -- (9,0.15); \draw[thick,-] (10,0) -- (10,0.15);

\node [green] at (2,0.15) {\large \textbullet}; \node [green] at (2,0.5) {\large \textbullet};\node [green] at (2,1.25) {\large \textbullet};\node [green] at (2,2.2) {\large \textbullet};\node [green] at (2,3.4) {\large \textbullet};\node [green] at (2,4.8) {\large \textbullet};

\node [green] at (0,0.9) {\large \textbullet};\node [green] at (4,0.9) {\large \textbullet};

\node [red] at (5,1.1) {\large \textbullet};  \node [red] at (9,1.1) {\large \textbullet};
\node [red] at (7,2.9) {\large \textbullet};  \node [red] at (7,4.5) {\large \textbullet};

\node [black] at (-0.05,-0.4) { $\frac{-\pi}{10}$ }; \node [black] at (2,-0.4) { $\frac{\pi}{10}$ }; \node [black] at (4,-0.4) { $\frac{3\pi}{10}$ };  \node [black] at (5,-0.4) { $\frac{2\pi}{5}$ };\node [black] at (7,-0.4) { $\frac{3\pi}{5}$ };\node [black] at (9,-0.4) { $\frac{4\pi}{5}$ };

  \end{tikzpicture}

\caption{A configuration of zeros of an eigenvalue of the transfer matrix corresponding to an excited state of the tricritical Ising model. Here we see the 1-strings of the first analyticity strip in green together with 1-strings and 2-strings of the second analyticity strips in red.}

\end{center} \end{figure}
\par\end{center}


\section{Conclusion}

In this paper, the $\mathcal{M}(5,6)$ relativistic integrable theory
was partially analyzed from the lattice point of view, in the $(r=1,\ s=1)$
sector. We described the patterns of zeros of the corresponding double
row transfer matrix eigenvalues and their $(m,n)$ systems. We adopted
a similar approach to analyze this model as was used before in \cite{BDeebP,Deeb,Deeb35}.
Other sectors of the boundary case are similar in their patterns of
zeros. The only difference is that some analytic strips would contain
a fixed zeroes at their centers.

Future work should extend the scope and exploit the lattice description
of the integrable scattering theory in order to fully solve the TBA
equations of the system and determine the spectrum of the model. The massive $\mathcal{M}(3,5)$
and $\mathcal{M}(5,6)$ models must be studied in following papers.
It also remains essential to study the same models using the bootstrap
methods.


\bibliographystyle{dcu}
\bibliography{ReferenceListRSOS}


\end{document}